\documentclass[aps,prl,preprint,superscriptaddress,floatfix]{revtex4-1}
\newcommand{\figSize}{.8}

\usepackage[english]{babel}
\usepackage{graphicx}
\usepackage{natbib}
\usepackage{color}
\usepackage{siunitx}

\makeatletter
\@ifundefined{textcolor}{}
{%
 \definecolor{BLACK}{gray}{0}
 \definecolor{WHITE}{gray}{1}
 \definecolor{RED}{rgb}{1,0,0}
 \definecolor{GREEN}{rgb}{0,1,0}
 \definecolor{BLUE}{rgb}{0,0,1}
 \definecolor{CYAN}{cmyk}{1,0,0,0}
 \definecolor{MAGENTA}{cmyk}{0,1,0,0}
 \definecolor{YELLOW}{cmyk}{0,0,1,0}
 }

\begin{document}
\title{A polariton electric field sensor}
\author{Emre Togan}
\affiliation{Institute of Quantum Electronics, ETH Zurich, CH-8093 Zurich, Switzerland}

\author{Yufan Li}
\affiliation{Institute of Quantum Electronics, ETH Zurich, CH-8093 Zurich, Switzerland}

\author{Stefan Faelt}
\affiliation{Institute of Quantum Electronics, ETH Zurich, CH-8093 Zurich, Switzerland}
\affiliation{Solid State Physics Laboratory, ETH Zurich, CH-8093 Zurich, Switzerland}

\author{Werner Wegscheider}
\affiliation{Solid State Physics Laboratory, ETH Zurich, CH-8093 Zurich, Switzerland}

\author{Atac Imamoglu}
\affiliation{Institute of Quantum Electronics, ETH Zurich, CH-8093 Zurich, Switzerland}

\begin{abstract}
We experimentally demonstrate a dipolar polariton based electric field sensor. We tune and optimize the sensitivity of the sensor by varying the dipole moment of polaritons. We show polariton interactions play an important role in determining the conditions for optimal electric field sensing, and achieve a sensitivity of 0.12 \si{V.m^{-1}.Hz^{-0.5}}. Finally we apply the sensor to illustrate that excitation of polaritons modify the electric field in a spatial region much larger than the optical excitation spot. 
\end{abstract}

\maketitle

Due to their interacting nature, microcavity polaritons have been extensively researched in the context of quantum fluids of light\cite{carusotto_quantum_2013}. Polaritons have also been utilized to demonstrate novel optoelectronic devices such as parametric amplifiers\cite{savvidis_angle-resonant_2000}, all optical transistors\cite{ballarini_all-optical_2013,gao_polariton_2012}, resonant tunneling diodes\cite{nguyen_realization_2013}, and all optical phase shifters\cite{sturm_all-optical_2014}. In an alternate application, we demonstrate that polaritons with a large permanent electric dipole moment (dipolar polaritons) are also sensitive electric field detectors. The detection principle relies on changes in the reflected optical power of a resonant laser upon the application of an electric field, which changes the transition energy of the dipolar polaritons. The resulting sensor, in principle, is fast, responds to changes in the electric field within the polariton lifetime, its sensitivity is tunable by changing the dipole moment of polaritons, and its ultimate performance is determined by the interactions among polaritons. Such a sensor is suited to optical detection of local electric fields that are within the sample. These electric fields could be created by other optically accessible excitations (e.g. indirect excitons\cite{stern_exciton_2014,high_spontaneous_2012}) or could be due to excitations that are optically inaccessible, for example dark excitons\cite{mazuz-harpaz_dynamical_2019}, and electric charges. The sensitivities that we achieve are sufficient to detect such particles far from the optical probing spot.

We realize polaritons that have an electric dipole moment by non-perturbatively coupling cavity photons with direct excitons (DX) and indirect excitons (IX) in an In$_{0.04}$Ga$_{0.96}$As coupled quantum well (QW) structure\cite{cristofolini_coupling_2012,togan_enhanced_2018}. Electron tunneling between the two QWs facilitates coherent coupling between IX and DX states. By embedding such structures in a planar cavity dipolar polaritons are formed. We focus on the lowest energy polaritons (lower polariton, LP) for sensing, and thus take advantage of the large electric dipole of indirect excitons (dipole length $d \sim 29$ nm) as well as the narrow linewidth ($\Gamma \sim 50$ \si{\micro eV}, FWHM) and the high optical reflection contrast of cavity photons. The change of the net growth direction electric field ($\Delta E$) at the polariton location leads to a change in the polariton transition energy $\sim e d \Delta E$, where $e$ is the electron charge, which in turn changes the reflected power of a laser that is nearly resonant with the polariton transition. We use this simple concept to realize the polariton electric field sensor.

\begin{figure}[hbtp]
\includegraphics[width= \figSize \textwidth]{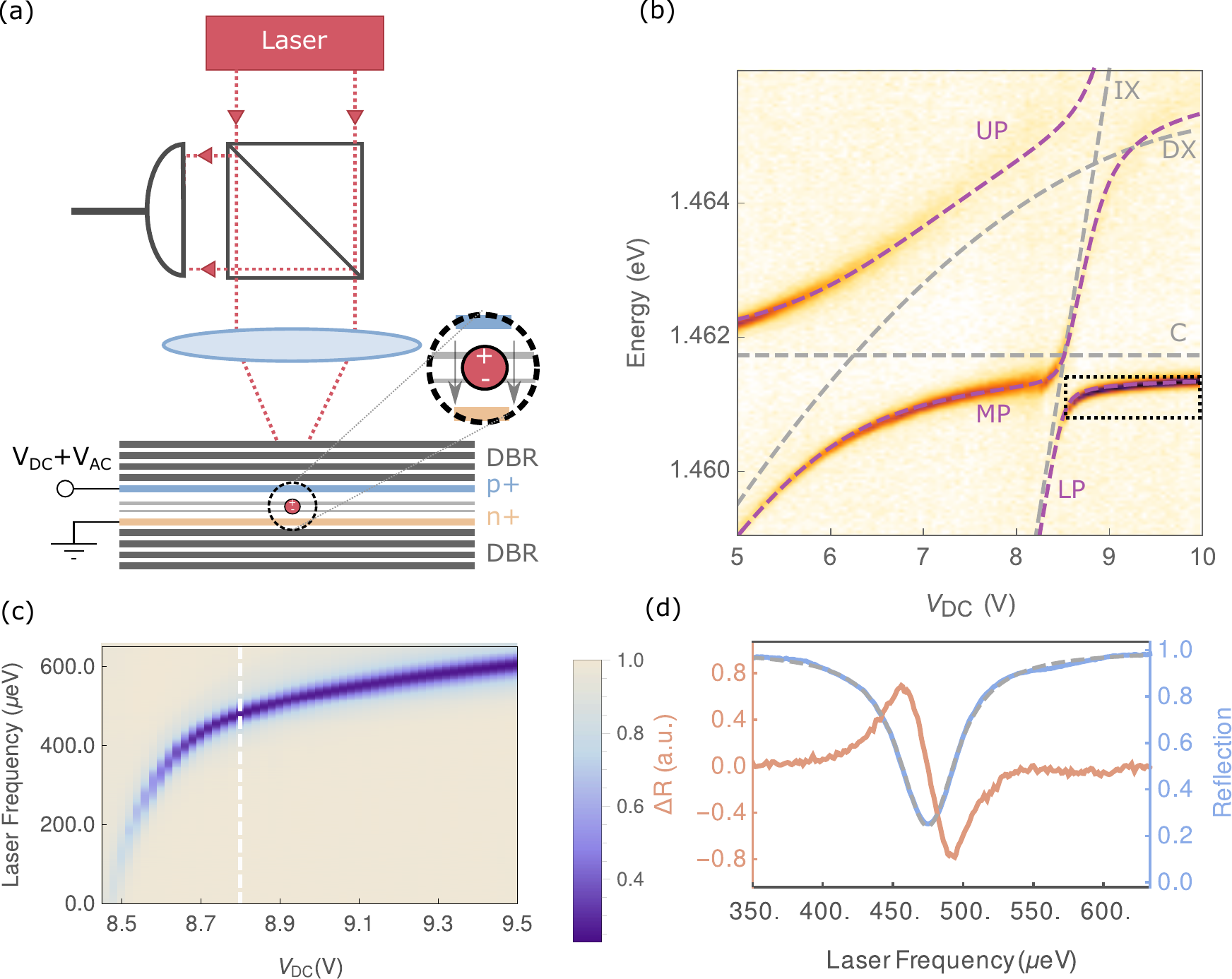}
\caption{ (a)  Schematic of the experimental setup and the sample structure. Sample is held at $\sim10$ K inside a flow cryostat. The coupled quantum well (QW) structure is located at an anti-node of a planar cavity that is formed between two distributed Bragg reflectors (DBRs).  We treat polaritons within the 10 \si{\micro m} optical excitation spot spot as a single mode. (b) White light reflection spectrum as a function of $V_{\mathrm{DC}}$. Three polariton branches: lower (LP), middle (MP) and upper (UP) are visible in the reflection spectrum (purple dashed lines) and the inferred energies of the cavity mode (C), IX and DX transitions are shown as gray dashed lines. (c) Laser reflection spectrum for the LP as a function of $V_{\mathrm{DC}}$ with 200 nW laser power. The vertical axis shows the laser frequency; 1.45989 eV shown as 0 eV. (d) In blue (right axis), line cut of (c) at $V_{\mathrm{DC}}= 8.78$ V.  In red (left axis) $\Delta R$, amplitude of the reflected power oscillating at the same frequency (76.226 kHz, 2 \si{mV}p-p) as the sub-linewidth modulation.
}
\label{fig1}
\end{figure}

The sample structure used in this Letter is illustrated in Figure~\ref{fig1}(a). The coupled QW structure resides in a $p\mathrm{-}i\mathrm{-}n$ diode, and we bias the diode with a DC potential ($V_{\mathrm{DC}}$) as well as a small AC potential ($V_{\mathrm{AC}}$). The diode bias changes the relative detunings between the different exciton levels, hence with $V_{\mathrm{DC}}$ we can tune the dipole moment, $ed$, of the polariton resonances.  Figure~\ref{fig1}(c) illustrates this for the LP, where the dipole length of the LP transition is tuned from $d = 0$ nm at $V_{\mathrm{DC}}=9.9$ V to $d = 2$ nm at $V_{\mathrm{DC}} = 8.5$ V. We use a sinusoidal AC potential to create a field to be sensed ($E_{\mathrm{AC}}$). The resulting controlled sub-linewidth  ($edE_\mathrm{AC} \ll \Gamma$) oscillating electric field is used to characterize performance and limits of the polariton electric field sensor. The oscillating field allows us to reduce the effects of low frequency noise that is present in our setup. A plot of the amplitude of the change in the reflected optical power at the modulation frequency (that we call $\Delta R$) as a function of laser frequency is shown in Figure~\ref{fig1}(d). As expected, $\Delta R$ signal is maximal at the points where the reflection signal has the highest derivative with respect to the laser frequency. We next characterize the sensitivity of the detector.

\begin{figure}[hbtp]
\includegraphics[width= \figSize \textwidth]{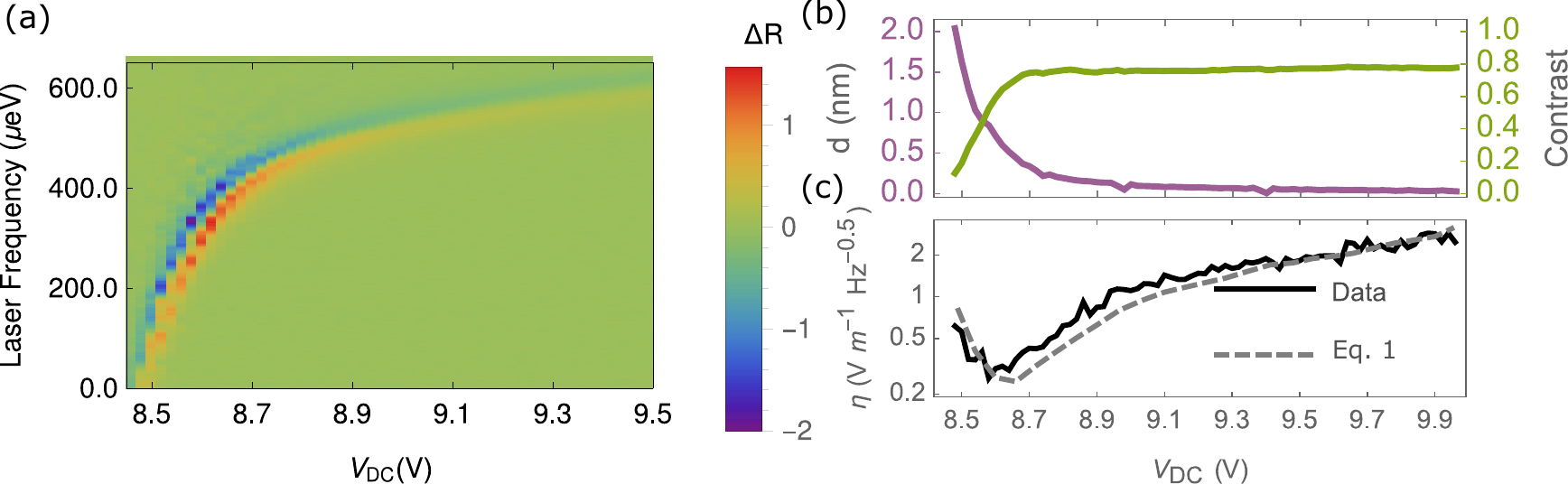}
\caption{ (a) Plot of $\Delta R$ (a.u.) as a function of laser frequency (1.45989 eV shown as 0 eV) and $V_{\mathrm{DC}}$ under same conditions as Figure~\ref{fig1}(c)-(d). (b) Dipole moment (left axis, purple line) and reflection contrast (right axis, green line) extracted from Figure~\ref{fig1}(c). The reflection contrast decreases as the dipole moment of the LP mode increases. (c) Experimental sensitivity as a function of $V_{\mathrm{DC}}$. For each value of $V_{\mathrm{DC}}$ we pick the detuning that yields the best sensitivity. For evaluation of Eq.~\ref{eq:sens} we use the experimentally determined values for reflection contrast and dipole moment shown in (b), the linewidth extracted from Figure~\ref{fig1}(c) and detected power $P$ in photons per second.
}
\label{fig2}
\end{figure}

The sensitivity ($\eta$) is defined as the minimum detectable field that yields a signal-to-noise ratio of unity for 1 second of integration~\cite{degen_quantum_2017}. To estimate the sensitivity we record a time trace of $\Delta R$ at each laser detuning for a duration of 1 ms. This 1 ms interval is broken in to sub intervals of length $\tau = 50$ \si{\micro s} (integration time). We determine $\Delta R$ for each of these sub intervals and record the value. We then calculate the standard deviation of $\Delta R$ within the 1 ms duration which gives us a signal to noise ratio for $\tau = 50$ \si{\micro s}. We use the fact that standard deviation of $\Delta R$ scales as $\tau^{-1/2}$ (see the Supplemental Material) to find the the signal-to-noise ratio for $\tau=1$ s. We use this inferred signal to noise ratio to determine the experimental sensitivity ($\eta$). The experimental sensitivity for the dataset shown in Figure~\ref{fig1}(c) is illustrated in Figure~\ref{fig2}(c). Figure~\ref{fig2}(a)\&(c) illustrates the maximum $|\Delta R|$ and the best sensitivity occurs at the optimal $V_{\mathrm{DC}}=8.6 $ V.

To better understand the existence of an optimum $V_{\mathrm{DC}}$ for the sensitivity, we note that for a shot noise limited linear sensor with a Lorentzian lineshape the sensitivity is given by (see Supplemental Material):
\begin{eqnarray}
\eta = \frac{\Gamma \sqrt{1-C/2}}{C e d \sqrt{P}}
\label{eq:sens}
\end{eqnarray}
where $\Gamma$ is the linewidth, $C$ is the reflection contrast, $P$ is the optical power in photons/s and we assumed a laser -- LP detuning of $\pm \Gamma/2$. All of the parameters in this expression are measured independently for each $V_{\mathrm{DC}}$ and we plot the expected sensitivity (with no additional fit parameters) along with the the experimental sensitivity in Figure~\ref{fig2}(c). The two are in close agreement and they both exhibit an optimum sensitivity around $V_{\mathrm{DC}} = 8.6$ V. The optimum in the sensitivity is due to the trade off between the increase in the dipole moment and the decrease in contrast as illustrated in Figure~\ref{fig2}(b). The increase in the dipole moment necessitates an increase in the exciton content of polaritons, that introduces non-radiative decay channels~\cite{delteil_towards_2019,munoz-matutano_emergence_2019}, which reduces the reflection contrast and increases the polariton linewidth. Changes in $V_{\mathrm{DC}}$ also lead to measurable, yet small, changes in the polariton linewidth (see Supplemental Material). Eq~\ref{eq:sens} predicts better sensitivites than what is measured: this difference is due to the additional noise from the detector, effects of slow variations in the resonance condition (see Supplemental Material), and laser induced electric field noise. We come back to the last point later in the text. This optimum is the central result of our work and it shows both the dipolar character and high reflection contrast is necessary to achieve good sensitivites. In particular at the optimum point with 200 nW of incident power, an electric field sensitivity of 0.26 \si{V.m^{-1}.Hz^{-0.5}} is achieved. 

\begin{figure}[htbp]
\includegraphics[width= \figSize \textwidth]{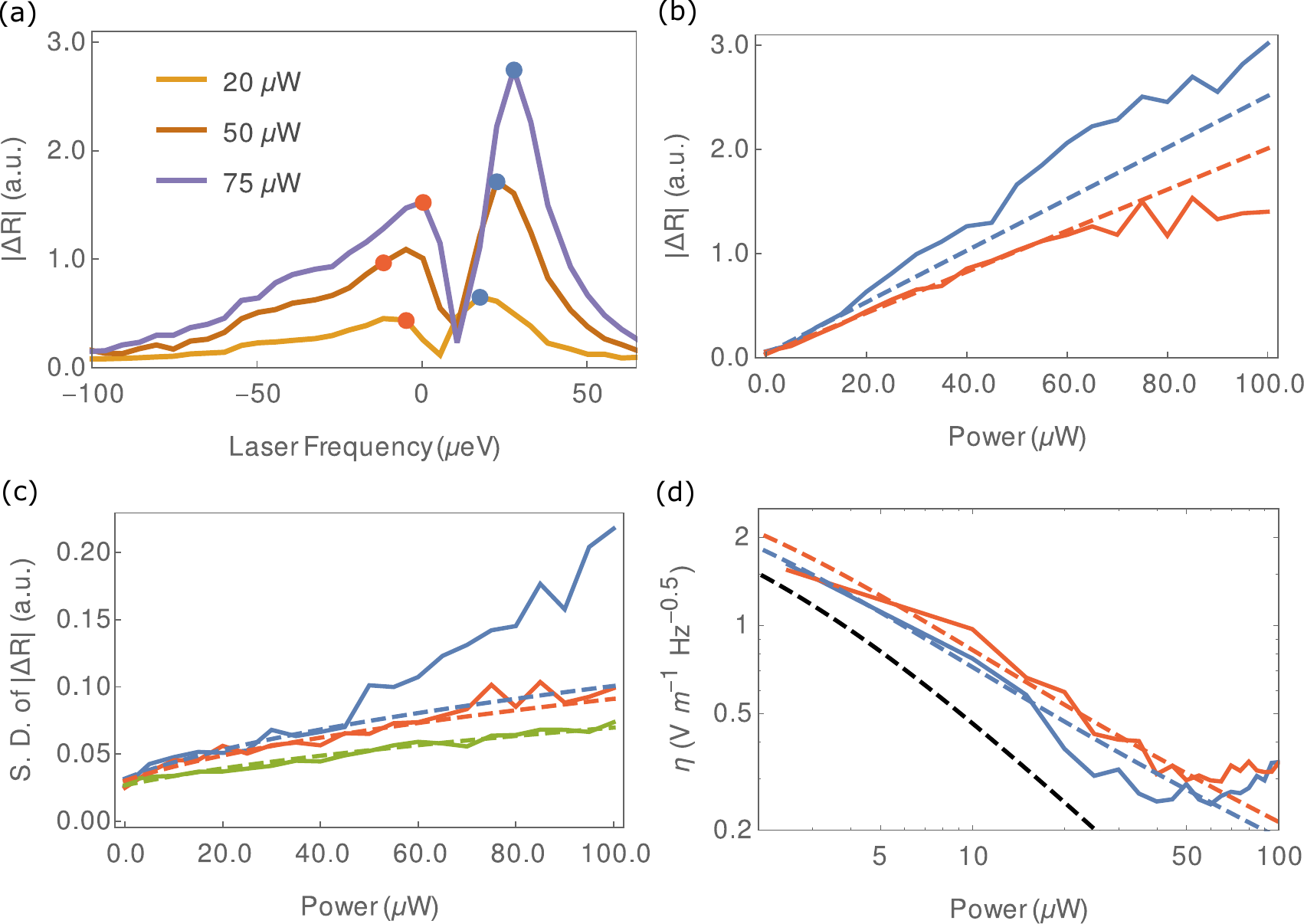}
\caption{ (a) Change of $|\Delta R|$ vs laser frequency (1.46065 eV shown as 0 eV) for different laser powers. Measurements performed  for 1.4 ms intervals with $\tau = 50$ \si{\micro s}. The solid dots indicate the frequency with the best sensitivity at a given power for blue and red detunings. (b) Plot of $|\Delta R|$ as a function of laser power at the laser frequency that yields the best sensitivity for each laser power. Red (blue) lines indicate red (blue) detuning relative to the polariton resonance. Linear fits to data $\leq 20$ \si{\micro W} are shown as dashed lines. (c) Standard deviation of $\Delta R$ as a function of laser power. The blue and red lines are for the same laser frequencies as in (b). Green line is measured at a relative laser frequency of -180 \si{\micro eV} (far red detuned). The dashed lines are a fit to $\sqrt{A^2+B^2 P}$ the expected standard deviation for detector noise and shot noise. For red and blue curves we fit data points $\leq  40$ \si{\micro W}. (d) Measured sensitivities as a function of laser power. Red and blue solid lines are experimental data measured at the same optimal laser frequencies as in (b). Red and blue dashed lines are the expected sensitivity based on the low power behaviour, shown as dashed lines in (b) and (c). Black dashed line is our best-case estimate for detector and shot noise limited sensitivity. }
\label{fig3}
\end{figure}

We next explore the effect of incident optical power on our sensor. To better illustrate the effects of higher optical power, we choose a higher modulation frequency (1.750181 MHz), a different gain setting on the photodiode, and $V_{\mathrm{DC}} = 9.8 $ V with weaker polariton interactions. These changes lead to an initial worse sensitivity for Figure 3 as compared to Figure 2. Figure~\ref{fig3}(a) shows $|\Delta R|$ as a function of laser frequency for three powers. At higher powers the polariton resonance blue shifts, and the lineshape becomes non-lorentzian~\cite{baas_optical_2004, fink_signatures_2018, togan_enhanced_2018}. Because of the blueshift, the frequency that yields the best sensitivity changes as a function of power. In addition, at higher powers, an asymmetry in the magnitude of $|\Delta R|$ develops between the blue and red sides of the resonance. As shown in Figure~\ref{fig3}(b), $|\Delta R|$ at the optimal frequency for sensitivity at a given power, increases super-linearly on the blue side, whereas it increases sub-linearly on the red side. The green dashed line in Figure~\ref{fig3}(c) illustrates a typical plot of the intensity noise as a function of power in the setup: it is measured as the standard deviation of $\Delta R$ at the laser frequency of $-180 \: \si{\micro eV}$ (far red detuned). It shows the typical power scaling expected of shot noise with sizeable detector noise. The measured noise at laser frequencies that yield the best sensitivities are above this noise level for powers $> 1 \: \si{\micro W}$. At powers higher than $40\:  \si{\micro W}$, the blue detunings exhibit more noise compared to red detunings.

Combination of these observations imply that the sensitivity of the polariton sensor deviates from the $\propto P ^{-0.5}$ power scaling in Eq.~\ref{eq:sens}. The major difference is the saturation of the sensitivity at high powers, observable in Figure~\ref{fig3}(d) for both blue and red detunings above $50\: \si{\micro W}$. The second difference is, for blue detunings, the sensitivity is better compared to the prediction of the linear model based on the low power behaviour (shown in blue dashed lines in Figures~\ref{fig3}(b)-(d)) for the power range 20 - 50 \si{\micro W}. We do emphasize that, even with this improvement, the sensor performs poorer compared to a best case sensitivity estimate shown as the black dashed line. This base case estimate is calculated with standard deviation estimated from the green dashed curve on Figure~\ref{fig3}(c) and $|\Delta R|$ estimated from the blue dashed curve on Figure~\ref{fig3}(b). 

At high powers, such as those used in Figure~\ref{fig3}, polariton interactions lead to a number of rich, nonlinear, phenomena that lead to deviations from the linear model used to derive Eq.~\ref{eq:sens}. 
The power range used in Figure~\ref{fig3} is slightly below the threshold for reaching one such phenomenon, optical bistability\cite{baas_optical_2004, boulier_polariton-generated_2014}. Within this power range we expect drastic changes in the lineshape, as well as significant changes in the intensity noise of the reflected light\cite{karr_squeezing_2004,boulier_polariton-generated_2014,fink_signatures_2018}. In terms of electric field sensing, changes in the lineshape lead to the super-linear (and sub-linear) increases of $|\Delta R|$ with power, and increased fluctuations in intensity of the detected light lead to increase in the standard deviation of $\Delta R$ and the eventual saturation of sensitivity with power. Ultimately the interactions limit the power range that can be utilized with polaritons, hence the sensitivity that can be achieved. In the results presented in Figure~\ref{fig3} there is significant noise in excess of the fundamental noise limit imposed by polariton interactions. We next turn our attention to laser induced electric noise that is ubiquitous in structures such as our sample\cite{coulson_electrically_2013,togan_enhanced_2018}.

\begin{figure}[hbtp]
\includegraphics[width= \figSize \textwidth]{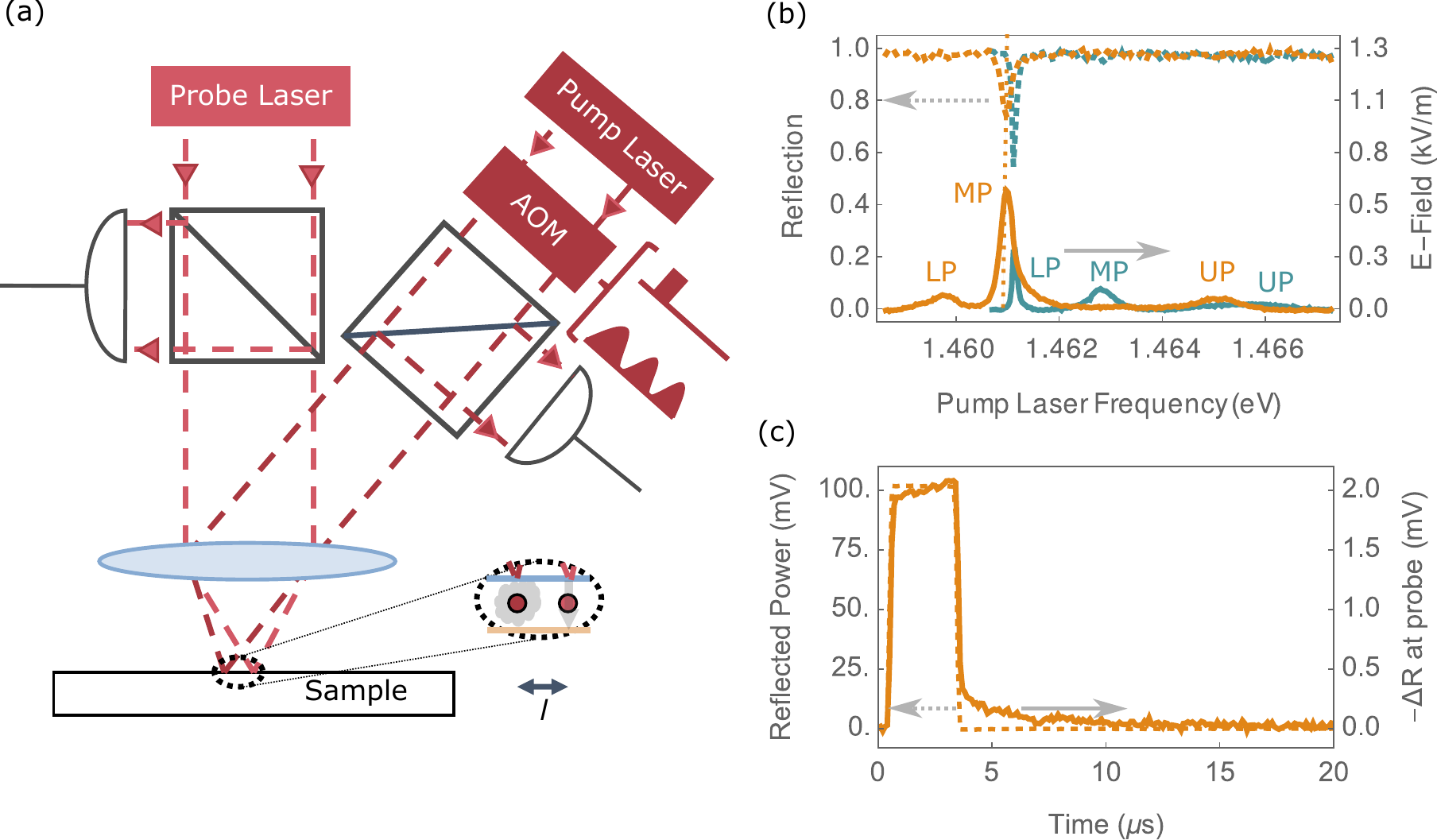}
\caption{ (a) Schematic of the setup used in the pump-probe (two-laser) experiments. We apply $V_{\mathrm{DC}} = 8.8$ V and using a galvo mirror, the probe laser and pump laser are focused on two separate spots on the sample. An AOM is used to modulate pump laser's power either in sinusoidal waveform (at 76.226 kHz, 100\% modulation depth) or a rectangular (3 \si{\micro s}) pulse. The reflection at both the probe and pump spots is measured simultaneously. (b) Reflected power at the pump spot (dotted lines left axis) and detected electric field at the probe spot (solid lines right axis) as a function of pump laser frequency. We show data at the same probe spot for two pump spots with $l = 176$ \si{\micro m} (orange) and $l = 65$ \si{\micro m} (green). Mean pump power is 500 nW. Different polariton resonances are identified on the electric field plots. Dashed vertical line shows the energy used for the pulsed experiment shown in (c). (c) Time trace of the reflected power at the pump spot ($l = 176$ \si{\micro m}, dashed line left axis) and detected electric field (solid line right axis) for the pump laser that is modulated to a rectangular pulse shape (3 \si{\micro s} long). 
}
\label{fig4}
\end{figure}

We utilize the polariton sensor to demonstrate an unexpected property of the laser induced electric field, namely that it has an influence on a length scale that is much larger than the excitation spot. We pick two spots on our sample that are separated by $l$. Polaritons on one spot (probe) are used as electric field sensors. We vary the conditions for exciting polaritons at a second (pump) spot and record changes in the sensed electric field (at probe) as a function of these conditions. The schematic of the setup used for this measurements is shown in Figure~\ref{fig4}(a).  On the probe spot, we excite polaritons with 200 \si{nW} of incident power and at a fixed laser detuning ($+\Gamma/2$). In a first experiment, we vary the photon frequency of the intensity modulated pump laser. A plot of the pump reflection and detected electric field, for $l = 60$ \si{\micro m} and $l = 176 $ \si{\micro m} as a function of the pump laser frequency is shown in Figure~\ref{fig4}(b). For both separations, the detected electric field shows three peaks that correspond to excitation of the three polariton branches at the pump spot. Remarkably the peak electric field detected is larger when $l$ is larger. At $l = 176 $ \si{\micro m} the excitation of the MP and at $l = 60 $ \si{\micro m} the excitation of the LP branch create the largest detected electric field. Due to the cavity wedge that is present in the sample, the relative exciton / cavity content of the different branches differ at the two pump locations. The MP at $l = 176 $ \si{\micro m} and the LP at $l = 60 $ \si{\micro m} have the largest cavity contents and narrowest linewidths. Consequently driving these resonances with the most pronounced reflection spectra lead to the largest number of polaritons. Remarkably other polariton branches that are barely visible in the reflection spectrum are clearly visible in the detected electric field.

In a second experiment we modulate the pump laser to have a rectangular pulse shape, tune its energy to the MP resonance and measure the detected electric field at the probe location as a function of time. As Figure~\ref{fig4}(c) demonstrates the sensed electric field exhibits rise and fall times within \si{\micro s} timescales, which is much longer compared to the polariton lifetime.

Excitation of polaritons at the pump spot causes changes in the electric field environment at different length scales. Excitation of dipolar polaritons create localized dipole-like electric fields, however the observed peculiar spatial dependence rules out dipolar contribution as the dominant detected field. Similarly we rule out creation of localized charge distributions, say due to occasional ionization of polaritons\cite{togan_enhanced_2018}. We attribute the dominant sensed electric field to the changes in the electric potential on the $p$ and $n$ doped layers due to the photocurrent that flows upon excitation of polaritons (see the Supplemental Material). Concurrently, the finite rise and fall times measured in Figure~\ref{fig4}(c) are attributed to the finite capacitance in the sample. The exact mechanism that leads to the photocurrent, as well as how to mitigate its presence remains an open question. Since the change in the potential occurs over a wide spatial region, feedback techniques can be used to mitigate such effects, moreover the slow rise and fall times that are measured suggest pump-probe experiments that probe dynamics on polariton lifetime would be unaffected by the influence of the photocurrent.

Our experiments demonstrate that polaritons are useful electric field sensors. The sensitivities that we have demonstrated $0.26 \: \si{V.m^{-1}.Hz^{-0.5}}$ in Figure~\ref{fig2} and $0.12 \:  \si{V.m^{-1}.Hz^{-0.5}}$ shown in Supplemental Material are an order of magnitude better compared to other solid state optical electric field sensors such as room temperature nitrogen vacancy center ensembles ($10 \: \si{V.m^{-1}.Hz^{-0.5}}$)\cite{michl_robust_2019}, or cryogenic single quantum dot based electric field sensors (1-$5 \: \si{V.m^{-1}.Hz^{-0.5}}$) \cite{vamivakas_nanoscale_2011,cadeddu_electric-field_2017,arnold_cavity-enhanced_2014}. The sensitivity is about two orders of magnitude worse compared to state of the art electric field sensors such as single electron transistors \cite{yoo_scanning_1997,devoret_amplifying_2000}, electromechanical resonators \cite{cleland_nanometre-scale_1998,bunch_electromechanical_2007}, and quantum enhanced Rydberg atom ensembles\cite{facon_sensitive_2016} that have demonstrated electric field sensitivities on the order of $1 \: \si{mV.m^{-1}.Hz^{-0.5}}$. The fundamental sensitivity achievable by polariton sensors can be improved, for example, by using state of the art DBR cavities, by using larger sensing areas and higher optical powers that then reduce the effective nonlinearity in the system, or by taking advantage of polariton number squeezing\cite{karr_squeezing_2004, boulier_polariton-generated_2014}.

As demonstrated in this paper, such polariton based electric field sensors are particularly suited to detecting charge or electric field distributions that are within the $p\mathrm{-}i\mathrm{-}n$ diode structure, such as the presence of remote dipolar polaritons. With the sensitivity demonstrated in this work, the electric field created by $\sim 1500$, 2 nm dipoles $15 \: \si{\micro m}$ away should be detectible in 1 s. This opens up possibilities for non-destructive detection of polaritons, for example in pump-probe experiments, and can be used as probes for the intra-cavity squeezing of polaritons. Polariton, or dipolar exciton, based sensing layers may be embedded during fabrication in close proximity to detect non trivial charge distributions in 2DEGs or Van-der-Waals materials\cite{shimazaki_moire_2019,tang_wse2ws2_2019,regan_optical_2019}. Such optical measurements allow for probing \si{\micro m^2} scale areas with good spatial resolution that are inaccessible with global electrical measurements. Moreover the polaritons are sensitive to magnetic fields due to the diamagnetic and Zeeman shifts\cite{pietka_magnetic_2015,lim_electrically_2017}, and based on our results we estimate magnetic field sensitivities on the order of $100 \: \si{nT.Hz^{-0.5}}$ with $1 \: \si{\micro W}$ of incident power are achievable.

\begin{acknowledgments}
The Authors acknowledge insightful discussions with Hyang-Tag Lim \& Young-Wook Cho. This work is supported by NCCR QSIT and an ERC Advanced investigator grant (POLTDES).
\end{acknowledgments}

%

\end{document}